\documentclass[prl,twocolumn,longbibliography,superscriptaddress]{revtex4-1}
\usepackage{amssymb,amsmath}
\usepackage{xcolor}
\usepackage{soul}

\DeclareMathOperator{\Tr}{Tr}

\usepackage{float}
\usepackage{upgreek}

\usepackage[english]{babel}
\usepackage{graphicx}
\usepackage{verbatim}
\usepackage{hyperref}
\newcommand{\Nmol}{\mathcal{N}_m}
\begin{document}
\title{Organic polariton lasing and the weak- to strong-coupling crossover}
\author{Artem Strashko}
\affiliation{SUPA, School of Physics and Astronomy, University of St Andrews, St Andrews, KY16 9SS, United Kingdom}
\author{Peter Kirton}
\affiliation{Vienna Center for Quantum Science and Technology, Atominstitut, TU Wien, 1040 Vienna, Austria}
\author{Jonathan Keeling}
\affiliation{SUPA, School of Physics and Astronomy, University of St Andrews, St Andrews, KY16 9SS, United Kingdom}
\date{\today}

\begin{abstract}
  Following experimental realizations of room temperature polariton lasing
  with organic molecules, we present a microscopic model that allows us to
  explore the crossover from weak to strong matter-light coupling.  We
  consider a non-equilibrium Dicke-Holstein model, including both strong
  coupling to vibrational modes and strong matter-light coupling,
  providing the phase diagram of this model in the thermodynamic limit.  We
  discuss the mechanism of polariton lasing, uncovering a process of
  self-tuning, and identify the relation and distinction between regular
  dye lasers and organic polariton lasers.
\end{abstract}

\maketitle

Bose-Einstein statistics underpin both the thermal equilibrium phenomenon of Bose-Einstein condensation, and the non-equilibrium phenomenon of lasing.  Lying between these two extremes, there now exist several experimental platforms; in particular exciton-polaritons (quasiparticles resulting from  strong coupling between photons and excitons) in semiconductor microcavities at cryogenic temperatures~\cite{Deng199,kasprzak2006bose,Balili1007},  and photons in dye-filled microcavities at room temperature~\cite{klaers2010bose}.  Since these microcavities are imperfect, they are sources of coherent light (as is a laser), but differ in mechanism from  photon lasing~\cite{RevModPhys_Deng,RevModPhys_Carusotto}. Indeed,  polariton lasing does not need electronic inversion, and so it is often stated that it can provide coherent light sources with ultra-low thresholds.  Polariton lasing can also occur at room temperature in appropriate materials:  inorganic materials such as wide bandgap  semiconductors~\cite{christopoulos07,guillet2011polariton,PhysRevLett.110.196406} and two-dimensional materials~\cite{waldherr2018observation}, and the focus of this Letter, organic materials~\cite{kena2010room,plumhof2014room,Daskalakis14}.  

Polariton condensation in organic materials prompts interesting questions regarding the mechanisms of polariton relaxation and lasing. Excitons in organic materials are Frenkel excitons --- electronic excitations of a molecule or chromaphore delocalized by hopping~\cite{agranovich2009excitations}.  Excitons in organic materials typically show complex absorption and emission spectra, due to strong coupling between the electronic state and the nuclear configuration, leading to rovibrational dressing. This causes a Stokes shift, so that emission is at longer wavelengths than absorption.  Spectral separation of emission and absorption underpins the operation of dye lasers~\cite{Schafer1990}, allowing gain without electronic inversion.  Since both strong matter-light coupling and large Stokes shifts are expected to reduce the lasing threshold, how they act in concert is of both fundamental interest and practical relevance.


Theoretical modeling of polariton condensates can follow a number of approaches.  To describe the macroscopic pattern formation and superfluid hydrodynamics it mostly suffices to use the phenomenological complex Gross-Pitaevskii equation~\cite{aranson2002world,RevModPhys_Carusotto}.  However, such  order parameter equations are ubiquitous in non-equilibrium systems breaking $U(1)$ symmetry,  so similar equations also apply for a photon laser~\cite{staliunas2003transverse,Berloff2013}.  Such approaches are thus  not well suited to understanding the relation between polariton and photon lasing, or the particular properties of organic polaritons.  To answer such questions, a more relevant approach is to use kinetic equations for the population of polaritons and excitonic reservoir, with decay rates accounting for vibronically assisted processes~\cite{Michetti2009,Fontanesi2009,Mazza2009,Michetti15} relevant in organic materials.  This approach has been used to understand the onset of lasing in anthracene microcavities~\cite{mazza2013microscopic} and  J-aggregated dyes~\cite{coles2011vibrationally}.  It however assumes polariton modes are well defined,  so cannot access the weak- to strong-coupling crossover.  In  this Letter we will work from a microscopic Hamiltonian which can, in the appropriate limit, recover both the physics described in such a kinetic model, as well as that of a dye laser in weak coupling.

We consider the non-equilibrium Dicke-Holstein model, presented below, which describes many molecules (with vibrationally dressed electronic transitions) coupled to a common photon mode.   The equilibrium phase diagram of this model has been presented elsewhere~\cite{cwik14}.
Similar models have been used to understand the absorption and photoluminescence spectra of organic polaritons~\cite{cwik2016excitonic,herrera2017dark,herrera2017theory,zeb2017exact}, and to explore whether strong matter-light coupling affects chemical reactions~\cite{herrera2016cavity,feist2017polaritonic,ribeiro2018polariton}.
Despite this, the non-equilibrium dynamics of this model, when both photons and vibrations couple strongly to the electronic state, has not been explored.  If coupling to vibrations is weak, one can eliminate these using Bloch-Redfield-Wangsness theory~\cite{wangsness53,redfield55} to capture the relaxation dynamics. For weak  coupling to light,  rate equations are available~\cite{kirton2013nonequilibrium,phot_cond_theor_PRA,hesten18,radonjic2018interplay}, describing photon condensation. For a few molecules, one can use exact numerical methods, treating the vibrational modes as a non-Markovian dissipation process~\cite{del2018tensor,del2018tensorB}.  However, the thermodynamic  limit with many molecules and strong coupling remains a challenge,  which we address here.

In this Letter we present the phase diagram of the non-equilibrium Dicke-Holstein model, exploring the crossover from weak to strong matter-light coupling. 
We uncover how the mechanism underlying polariton lasing evolves with coupling strength. In the weak coupling limit, we recover the results of models of photon condensation~\cite{kirton2013nonequilibrium,phot_cond_theor_PRA}, while for strong coupling our results are consistent with kinetic models~\cite{Mazza2009}. Finally, we address the practically significant question of whether strong coupling reduces lasing threshold, and the optimal parameters to realize polariton lasing.  

\begin{figure}[ht]  
\includegraphics[width=0.99\linewidth]{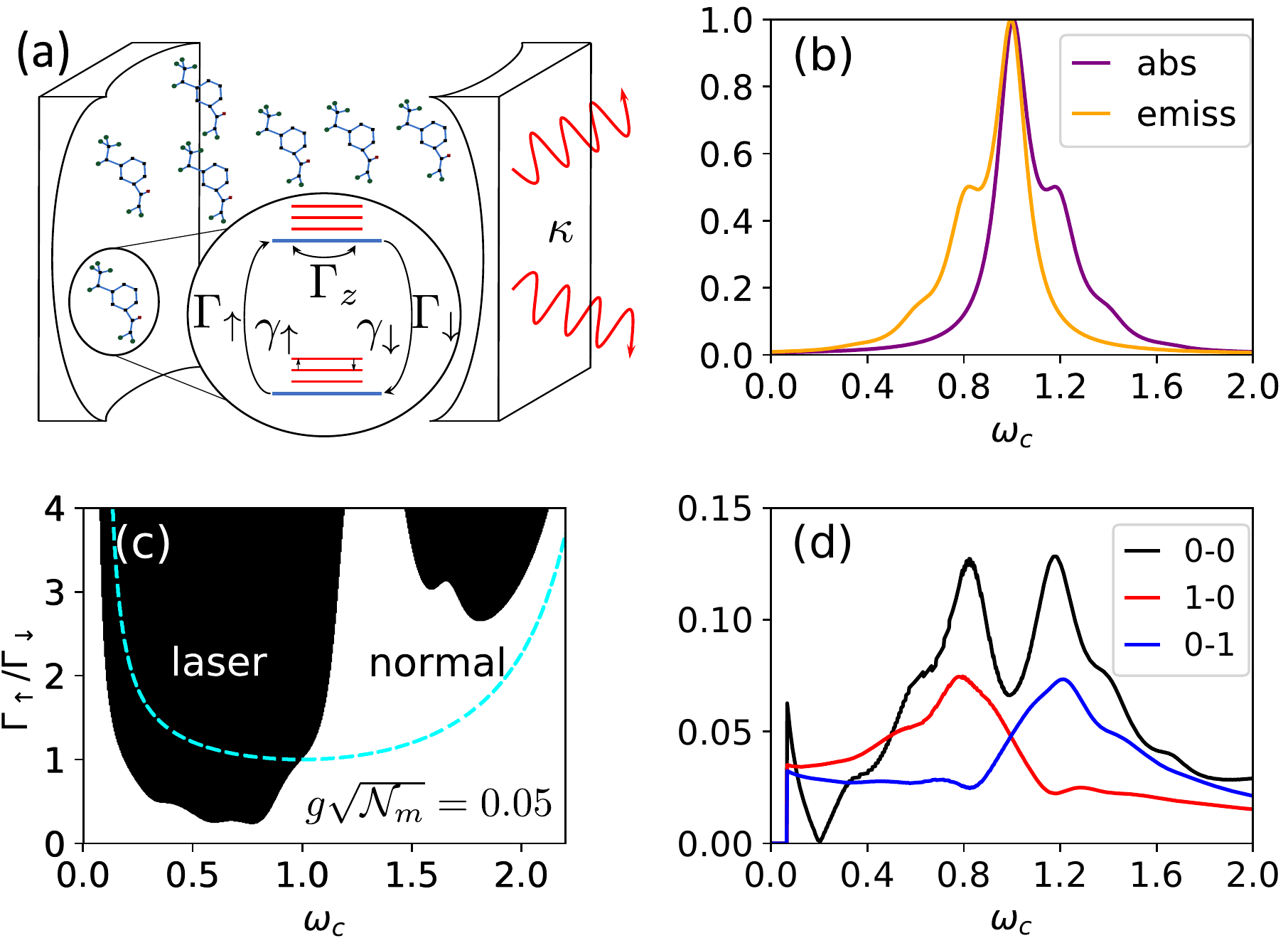}
\caption{(a) Cartoon illustrating our model: many molecules
  ($N$-level systems) are  coupled to a cavity mode. (b-d) Weak coupling
  behavior.  (b)  Emission and absorption spectra of the
  molecules. (c) Weak coupling phase diagram.  The cyan dashed line marks the phase boundary without vibrational dressing ($S=0$). (d)  Dominant molecular
  transitions coupled into the lasing mode at threshold.  All energies are
  measured in units of $\varepsilon\equiv 1$, other parameters are $S =
  0.1$, $\omega_v = 0.2$, $\Gamma_{\downarrow} = \kappa = 10^{-4}$,
  $\Gamma_z = 0.03$, $\gamma_{v} = 0.02$, $k_B T_v = 0.025$, $N_v=4$.}
\label{cartoon}
\end{figure}

We consider a single cavity mode, with frequency $\omega_c$, coupled to $\Nmol$ organic molecules as illustrated in Fig.~\ref{cartoon}(a). We model each molecule as two electronic states (HOMO and LUMO levels), dressed by a bosonic mode describing an internal molecular vibrational mode.  This yields the Dicke-Holstein model:
 \begin{multline}
\label{origin}
H =
  \sum_n
  \left\{
    \varepsilon \sigma_n^z  + 
  \omega_v
  \left[
    b^{\dagger}_n b_n^{\mathstrut} +
    \sqrt{S} \sigma_n^z (b^{\dagger}_n + b_n^{\mathstrut})
  \right]
  \right\}  \\ + 
    \omega_c a^{\dagger} a + 
    g \left( a^{\dagger} + a \right) \sum_n \sigma_n^x + 
    \frac{g^2 \Nmol}{\varepsilon} (a^{\dagger} + a)^2,
\end{multline}
where $\sigma_n^{\alpha=x,y,z}$ are Pauli matrices describing the electronic state of molecule $n$, $b^{\dagger}_n$ creates a vibrational excitation on molecule $n$, and $a^{\dagger}$ creates a photon.  To explore the ultrastrong matter-light coupling regime we do not make a
 rotating-wave approximation and we include a diamagnetic $A^2$ term, which prevents a ground state superradiant transition  (see~\cite{hepp1973equilibrium,rzazewski1975phase,griesser16,de2018cavity,Supp}). 
We note the diamagnetic term could be eliminated by a Bogoliubov
transform, yielding a model without the $A^2$ term, but with a modified photon frequency ${\omega_c^{\text{eff}} = \sqrt{\omega_c \left( \omega_c + 4 g^2 \Nmol / \varepsilon \right)}}$, and matter-light coupling $g^{\text{eff}} = g \sqrt{\omega_c/\omega^{\text{eff}}_c}$.
  We characterize matter-light coupling by the ``bare'' polariton splitting $g\sqrt{\Nmol}$, which would be the  splitting between the upper and lower polariton in the limit $S\to 0$.

To include incoherent pumping and decay processes, we use a Lindblad master equation~\cite{Breuer2002} of the form:
\begin{multline}
  \label{eq:orig_lindblad}
  \dot{\rho}(t) = 
  -i \left[ H, \rho \right] + \kappa \mathcal{L}[a] + 
  \sum_n \Big(
  \Gamma_{\downarrow}  \mathcal{L}[\sigma_n^{-}] +
  \Gamma_{\uparrow}  \mathcal{L}[\sigma_n^{+}] 
  \\  +
  \Gamma_z \mathcal{L}[\sigma_n^{z}] +
  \gamma_{\uparrow}  \mathcal{L}[b^{\dagger}_n - \sqrt{S} \sigma^z] +
  \gamma_{\downarrow}  \mathcal{L}[b^{\mathstrut}_n - \sqrt{S} \sigma^z]
  \Big),
\end{multline}
where $\mathcal{L}[X] =  X \rho X^{\dagger} - \{X^{\dagger} X, \rho\}/2$. 
We include electronic excitation and decay via non-cavity modes with rates $\Gamma_{\uparrow}, \Gamma_{\downarrow}$ respectively, dephasing with rate $\Gamma_z$, and photon loss with rate $\kappa$. 
The final two terms describe relaxation of the vibrational mode
to thermal equilibrium  at temperature $T_v$, accounting for the electronic-state-dependent vibrational
displacement.  These rates are thus
$\gamma_{\uparrow} = \gamma_v n_B$, $\gamma_{\downarrow} = \gamma_v (n_B + 1)$ where $n_B = [\exp( \omega_v/k_B T_v) -1]^{-1}$.
 Throughout this Letter we measure energies in units
of the electronic transition energy, so $\varepsilon\equiv 1$ by definition --- typical physical values are $\varepsilon \simeq 1$--$2$eV ---
other parameters are given in the caption of Fig.~\ref{cartoon}. We choose parameters that are physical and demonstrate sidebands 
in the molecular spectra, Fig.~\ref{cartoon}(b). 
The resulting weak coupling phase diagram,  Fig.~\ref{cartoon}(c), is straightforward to understand.   The threshold $\Gamma_{\uparrow}^{th}$  is reduced (i.e. $\Gamma_{\uparrow}^{th} < \Gamma_{\downarrow}$) when the cavity frequency matches the emission peak, and $\Gamma_{\uparrow}^{th}$ increases
when the cavity matches the absorption peak. 

Our aim is to consider the large $\Nmol$ limit of this model exactly, allowing both for strong matter-light coupling and strong coupling between vibrational and electronic states.  
In the large $\Nmol$ limit, a mean-field (i.e. Maxwell-Bloch) treatment of the Dicke model becomes exact~\cite{haken70,Dicke_MF_eqs}. There is however no  decoupling between electronic and vibrational degrees of freedom. 
Instead we define generalized molecular operators, describing an $N$-level system as a whole, with $N=2N_v$ for $N_v$ vibrational levels, where the choice of $N_v$ depends on the value of $S$.  We write the operators for this $N$-level system using a basis of generalized Gell Mann matrices $\lambda_i$~\cite{stone2009mathematics}, satisfying $\text{Tr}(\lambda_i \lambda_j) = 2 \delta_{ij}$.  This enables us to write any operator as
$O =  (\lambda_i/2)  \text{Tr} \left( O \lambda_i \right)$~\footnote{In writing this, we have defined $\lambda_0 =(2/N) \mathbf{1}_{N}$.}.
We may then write the Hamiltonian as:
\begin{equation}
\label{new_ham}
H =
  \omega_c a^{\dagger} a 
  +
  \sum_n
  \left[
    A_i + B_i ( a^{\dagger} + a )
  \right]
  \lambda_i^{(n)}
  + \frac{g^2 \Nmol}{\varepsilon} (a^{\dagger} + a)^2,
\end{equation}
with summation convention over $i$,
where $A_i, B_i$ can be found by constructing the molecular operators in
Eq.~(\ref{origin}) as $N\times N$ matrices, and taking traces.
The Lindblad master equation can be rewritten in the same way:
\begin{equation}
\label{new_Lindblad}
\dot{\rho}(t) =
-i \left[ H, \rho \right] +
  \kappa \mathcal{L}[a] +
  \sum_{\mu n} \mathcal{L}\left[ c_i^{\mu} \lambda_i^{(n)} \right].
\end{equation}
The sum over $\mu$ is over the five molecular dissipative channels
in Eq.~(\ref{eq:orig_lindblad}). If these terms are written as $\sum_\mu \Gamma_\mu \mathcal{L}[J_\mu]$  then we have $c_i^{\mu} = \sqrt{\Gamma_\mu} \text{Tr}(J_\mu \lambda_i)/2$.

The mean-field decoupling is  realized by 
deriving the equations of motion for variables
$\alpha = \langle a \rangle, \ell_i=\langle \lambda_i \rangle$, 
where $\langle O \rangle = \Tr ( \rho O )$, and then making the mean-field decoupling:
$\langle a \lambda_i\rangle = \langle a \rangle \langle \lambda_i \rangle$.
This leads to the set of nonlinear coupled differential equations:
\begin{align}
\partial_t \alpha &= 
  - \left( i \omega_c + \frac{\kappa}{2} \right) \alpha - 
  4 i \frac{g^2 \Nmol}{\varepsilon} \mathrm{Re}[\alpha] - 
  i \Nmol B_i \ell_i,
  \\
\partial_t \ell_i &= 
  \Big( \xi_{ik} + 4 f_{ijk} B_j \text{Re} [\alpha]  \Big) \ell_k + 
  \frac{4i}{N} c_j^{\mu} c_k^{\mu \ast} f_{ijk},
\end{align}
where $\xi_{ik}= 2 f_{ijk} A_j + i c_l^\mu c_m^{\mu \ast} (f_{ilp} \zeta_{mpk} + f_{mip} \zeta_{plk})$ with $\zeta_{ijk} \equiv \text{Tr}(\lambda_i \lambda_j \lambda_k)/2$, and 
$f_{ijk} \equiv \text{Tr}([\lambda_i ,\lambda_j] \lambda_k)/4i$.

Mean-field theory shows a phase transition between a normal state with $\alpha=0$, and a state with $\alpha \neq 0$ which we denote
as a laser but which 
 may be either photon or polariton lasing.  The phase boundary  can be found by considering when fluctuations about the normal state are unstable. 
 We thus write linearized equations of motion for fluctuations $\alpha = \delta \alpha$, $\ell_i = \ell_{i,\text{ns}} + \delta \ell_i$, where $\ell_{i,\text{ns}}$ is the normal state solution.  Defining $\upsilon = \left( \delta \alpha , \delta \alpha^* , \mathbf{\delta\ell} \right)^{\intercal}$ we can write ${ \partial_t \upsilon = \mathcal{M} \upsilon }$, 
and find the eigenmodes, $\mathcal{M} v^k = \xi^k v^k$. The  real  part of $\xi^k$ gives the growth (positive) or decay (negative) rate of a mode, while the imaginary part gives its oscillation frequency.

From the eigenvector $\upsilon^k$ we can also find the contributions of different molecular transitions to the given unstable mode (see~\cite{Supp} for details.)  As an example of this,  
Fig.~\ref{cartoon}(d) shows the composition of the unstable mode precisely at the lasing threshold, i.e.\ along the phase boundary shown in Fig.~\ref{cartoon}(c).  We see that where the threshold is low,  the (1--0) transition contributes most, while  where the threshold is high, the (0--1) transition dominates. In this Letter we show results with up to three vibrational excitations, 
i.e.\ $N=8$.  Our results are converged for this choice as shown in~\cite{Supp}.

\begin{figure}[ht]  
\includegraphics[width=0.99\linewidth]{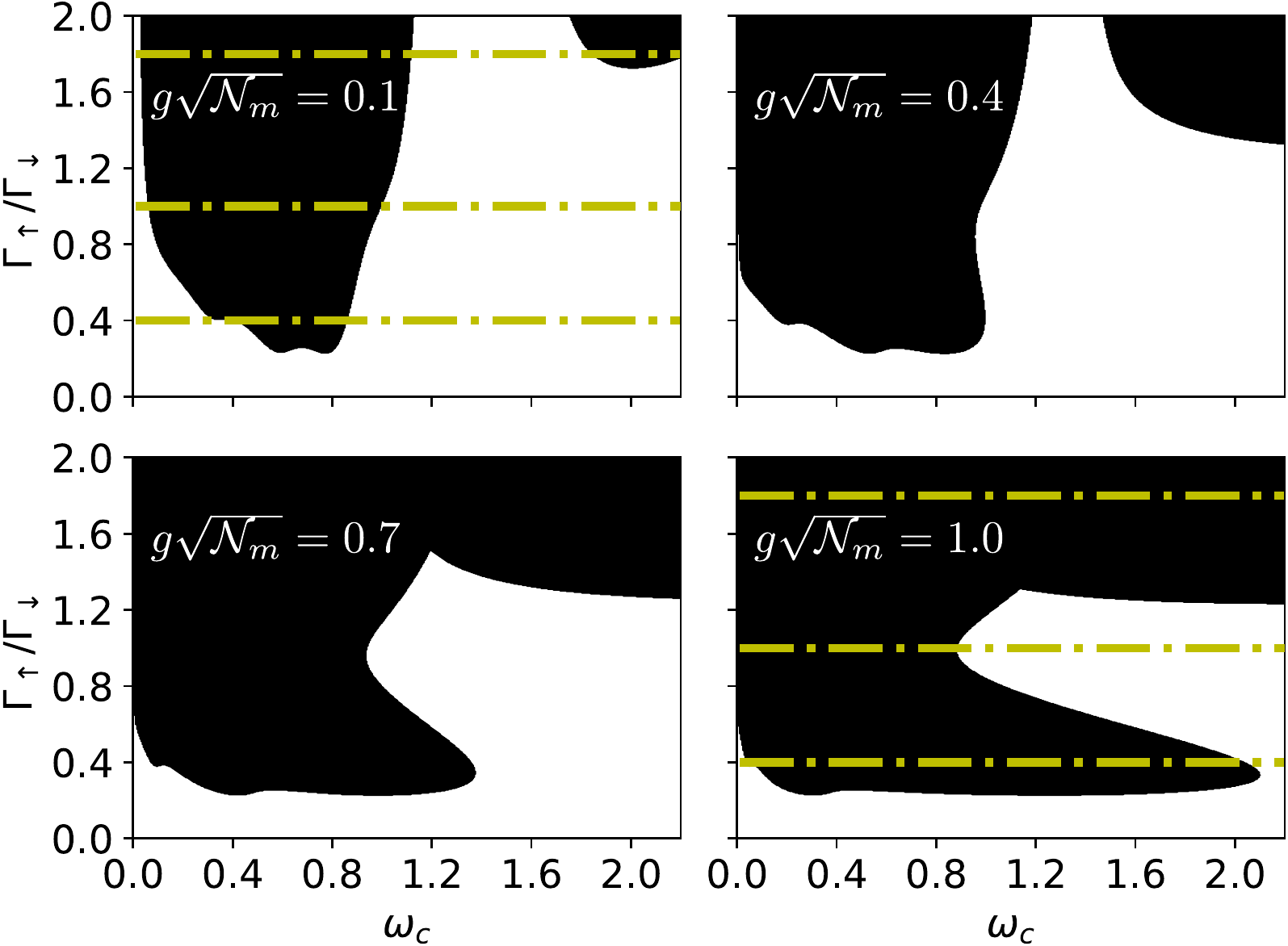}
\caption{Evolution of phase diagrams with increasing coupling, $g
  \sqrt{\Nmol}$ (values as shown).  
  Dash-dotted (yellow) lines indicate the cuts shown in
  Fig.~\ref{spectrum_and_DM_cuts}.  Parameters as in Fig.~\ref{cartoon}.}
\label{ph_diagrams_g_goes_up}
\end{figure}

Using the methods above, we now explore how strong coupling to light modifies the phase diagram, and understand the physics responsible for this modification.  Figure~\ref{ph_diagrams_g_goes_up} shows 
the evolution of the $\omega_c, \Gamma_\uparrow$ phase diagram as matter-light coupling increases, focusing on the low pumping regime. At moderate coupling, $g \sqrt{\Nmol} = 0.1$, we see the same form as for weak coupling, with a minimum (maximum) threshold at $\omega_c \simeq \varepsilon \mp \omega_v$.
As $g \sqrt{\Nmol}$ increases, the most striking feature is that the lobe at 
$\omega_c \approx \varepsilon - \omega_v$ bends and, at weak pumping, extends to significantly higher cavity frequencies. This eventually leads to
a re-entrant phase diagram.

\begin{figure*}[t]  
\includegraphics[width=0.99\linewidth]{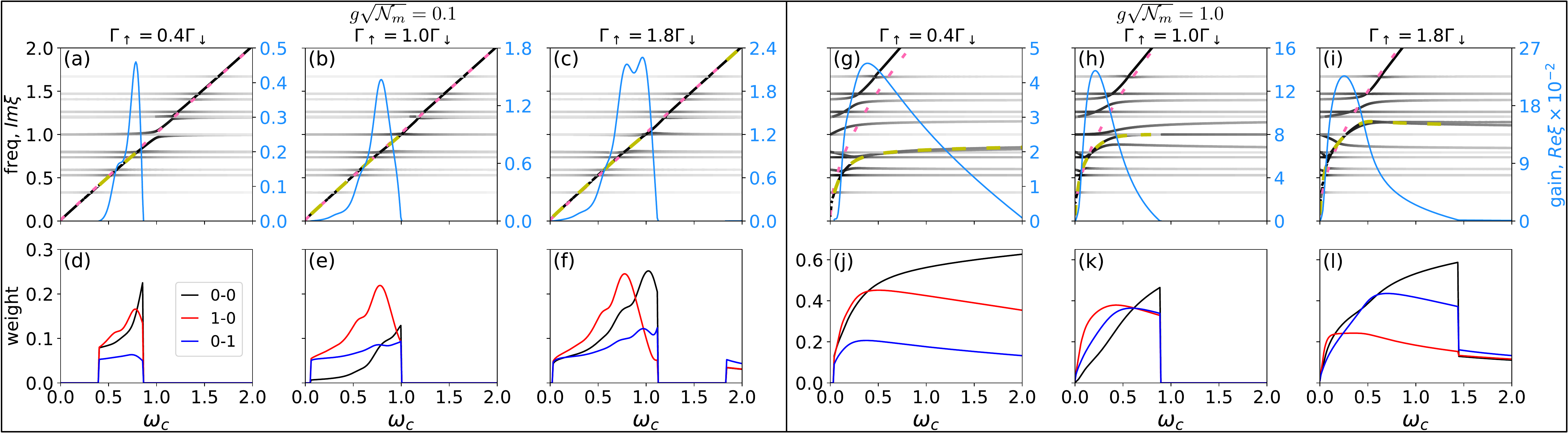}
\caption{Nature of lasing instability.  Top (a--c,g--i): Real (right, solid cyan) and imaginary (left, grayscale) parts of linearized eigenvalues $\xi$.  The grayscale
  indicates the photon component of that mode.  The yellow dashed line
  highlights which imaginary part corresponds to the mode that is unstable
  (i.e has a positive real part).  The pink dotted line shows the effective
  photon frequency, $\omega_c^{\text{eff}}$.  Bottom (d--f, j--l): Vibrational composition of unstable mode.   Parameters as in  Fig.~\ref{cartoon}. }
\label{spectrum_and_DM_cuts}
\end{figure*}

To understand the origin of this extension of the lobe to high cavity frequencies, we explore the nature of the unstable mode.
Figure~\ref{spectrum_and_DM_cuts} shows the composition of the unstable
mode, and the frequency of all modes for three cuts across the phase diagram  at fixed pump power at weaker and stronger coupling.  
For orientation, we first summarize the more straightforward behavior seen at weaker coupling, $g \sqrt{\Nmol} = 0.1$. In Fig.~\ref{spectrum_and_DM_cuts}(a), for low pumping $\Gamma_{\uparrow}  = 0.4 \Gamma_{\downarrow}$,  
the frequencies of the normal modes show a small polaritonic splitting  where the effective photon frequency~\footnote{{i.e.\ } this line plots the quantity
$\sqrt{ \omega_c \left( \omega_c + 4 g^2 \Nmol / \varepsilon \right)}$, hence its nonlinear dependence on $\omega_c$  at strong coupling} crosses molecular transitions, 
most clearly the zero phonon line at $\text{Im}[\xi] =  \varepsilon \equiv 1$,  i.e the (0--0) transition. Lasing occurs when the bare photon frequency is 
close to the (1-0) or (2-0) transitions (but with different strengths for these two transitions). Indeed, as seen in Fig.~\ref{spectrum_and_DM_cuts}(d), the unstable mode predominantly 
involves the (1--0) transition, crossing over to the (0--0) transition as $\omega_c$ approaches the zero phonon line. As pumping increases to $\Gamma_{\uparrow} = \Gamma_{\downarrow}$, 
Fig.~\ref{spectrum_and_DM_cuts}(b,e), saturation suppresses the polariton splitting, and lasing is possible over a wider range of frequencies. 
Finally, at $\Gamma_{\uparrow} = 1.8 \Gamma_{\downarrow}$, Fig.~\ref{spectrum_and_DM_cuts}(c,f),  we have electronic inversion, and can achieve lasing even when absorption exceeds emission.  
Here we have two lasing regions at low and high frequencies.
At yet higher pump strengths, these  regions join to form a single region.

At  stronger coupling, $g \sqrt{\Nmol} = 1.0$, the picture changes dramatically. For weak pumping, Fig.~\ref{spectrum_and_DM_cuts}(g) now shows a much clearer 
anticrossing, but also shows a new feature: locking between the frequencies of the polariton and of the (1--0) vibrational sideband.  This frequency 
locking persists over the range of cavity frequencies for which lasing occurs.  As such, although the bare  photon frequency is high, the lower polariton 
mode is at a lower frequency, and is self-tuned to allow feeding by the (1--0) molecular transition~\cite{mazza2013microscopic}.  Thus, at strong coupling, the system can self-tune to support such feeding.

As pumping increases to $\Gamma_{\uparrow} =  \Gamma_{\downarrow}$, the polaritonic self-tuning effect reduces and we no longer have lasing over such a wide range 
of cavity frequencies. As noted above, stronger pumping suppresses the polaritonic splitting, and this prevents the self-tuning effect.  
We nonetheless still have notable polaritonic splitting at low cavity frequencies. The  molecular transitions involved in lasing in Fig.~\ref{spectrum_and_DM_cuts}(k) is similar to that in Fig.~\ref{spectrum_and_DM_cuts}(e), but now (0--1) and (1--1) transitions play a larger role, as strong coupling admixes more electronic transitions into the polariton.
As we further increase pumping to $\Gamma_{\uparrow} = 1.8 \Gamma_{\downarrow}$ we now have two distinct lasing regimes: at low cavity frequency, 
there is polariton lasing, but at high cavity frequencies, lasing is at the bare cavity photon frequency (see~\cite{Supp} for an extended figure).

\begin{figure}[ht]  
\includegraphics[width=0.99\linewidth]{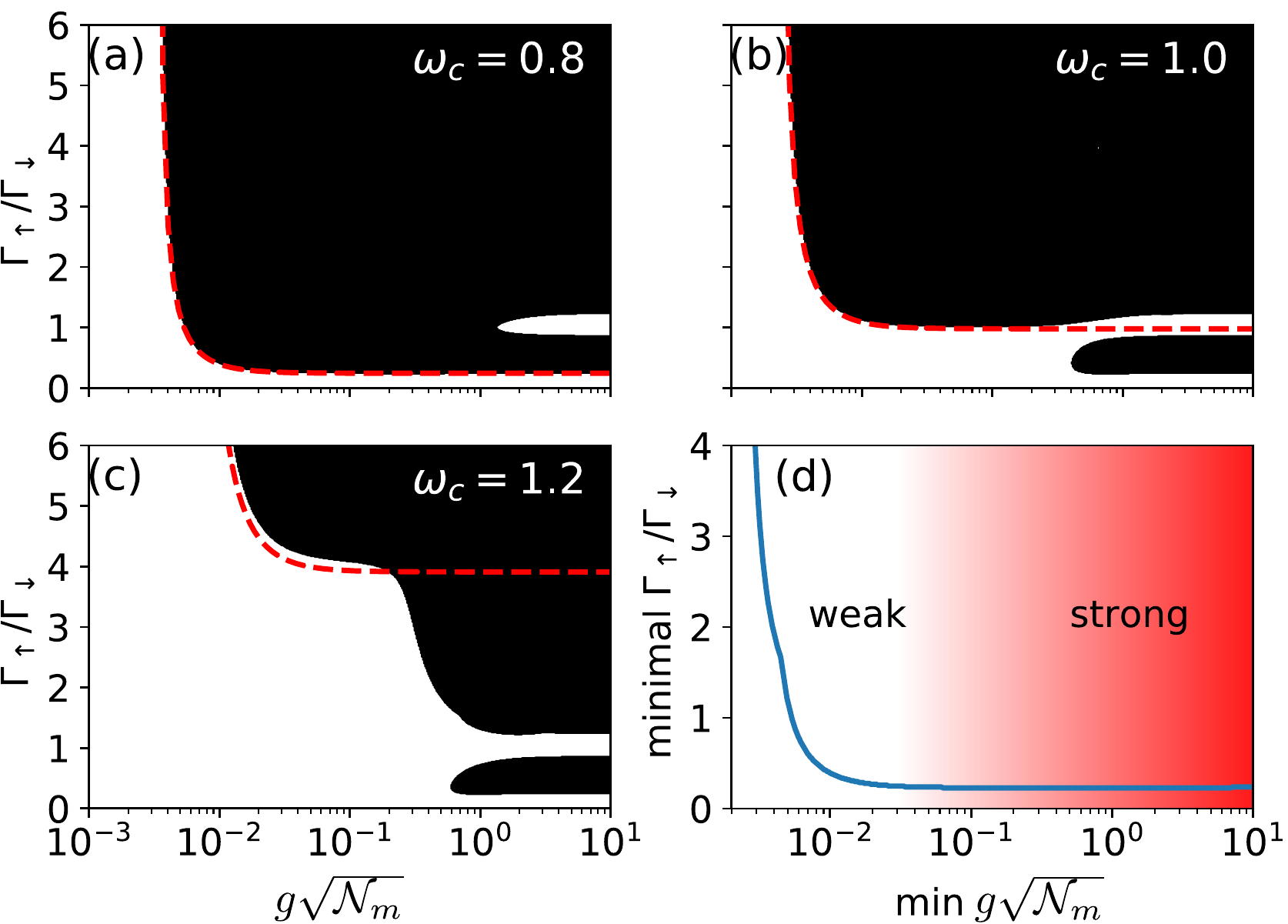}
\caption{(a-c) Phase diagrams at various cavity frequencies. Red dashed line shows the weak-coupling theory phase boundary, see~\cite{Supp}. 
(d)  Minimal critical pump strength (optimized over cavity frequency) as a function of matter-light coupling.  Red shading indicates weak- to strong-coupling crossover.}
\label{threshold}
\end{figure}

Having analyzed the structure of the phase diagram, we next consider whether strong matter-light coupling is a direct route to reduce the lasing threshold. 
In Fig.~\ref{threshold}(a--c) we plot phase diagrams vs $\Gamma_{\uparrow}$ and $g\sqrt{\Nmol}$ at three different bare photon frequencies, and we compare these to the phase boundary predicted by weak coupling theory, see~\cite{Supp} for details.  In all cases, we see the threshold reduce with increasing coupling. However,  for cavity frequencies near the (1-0) transition, Fig.~\ref{threshold}(a), the weak coupling prediction of the phase boundary matches the full results well.  
For this optimal frequency, while threshold pumping does reduce with increasing coupling, the threshold saturates as the system enters the strong coupling regime (where $g \sqrt{\Nmol}$ is larger than the linewidth $\Gamma_z$).
However, for cavity frequencies near the absorption peak at the
(0-1) transition, Fig.~\ref{threshold}(c), one sees the full calculation predicts a dramatically lower threshold than is predicted by weak coupling, and the threshold continues to reduce even after entering the strong coupling regime.

In Fig.~\ref{threshold}(d) we summarize the data by plotting the minimal pumping strength (optimized over cavity frequency~\cite{Supp}) required for lasing at a given matter-light coupling strength.  
While this optimized pump strength changes little above $g\sqrt{\Nmol} =0.1$, the results of Fig.~\ref{ph_diagrams_g_goes_up} clearly show that the range of cavity frequencies compatible with low threshold lasing 
does increase significantly.

In conclusion, we have studied the phase diagram of the non-equilibrium Dicke-Holstein model.  We showed that at strong coupling,  self-tuning of the polariton to optimize feeding by a vibrational sideband leads to lasing over a wide range of cavity frequencies.  We also see that the minimum achievable threshold is reduced as one approaches strong coupling, but does not change  as coupling is further increased.
Our results and approach also open up a number of possible future directions to explore.
For example, one may go beyond mean-field description with a cumulant 
expansion~\cite{meiser09,meiser10,Dicke_MF_eqs,kirton2018superradiant} and explore the spectral properties of emission and also luminescence below lasing threshold.
Further, by considering a multimode Dicke model, one can ask about the thermalization of mode populations, analogous to that seen in the weak-coupling  photon 
BEC~\cite{klaers2010bose,kirton2013nonequilibrium,phot_cond_theor_PRA,marelic15,hesten18,radonjic2018interplay}. 
Finally, one may consider more complex models, for example combining multiple vibrational modes in the system, or using Redfield theory with the true system eigenstates to better capture the physics of structured baths, 
leading to a more realistic master equation~\cite{ciuti05,del2015quantum}.  This may allow one to capture physics that otherwise requires numerically intensive non-Markovian 
simulations~\cite{del2018tensor,del2018tensorB}. 
Taken together, these provide an efficient route to understand the full range of weak and strong matter-light coupling for polariton 
condensation and lasing.

\begin{acknowledgments}
  AS acknowledges support from the EPSRC CM-CDT (EP/L015110/1).
  JK and PK acknowledges financial support from EPSRC program "Hybrid
  Polaritonics" (EP/M025330/1). PK acknowledges support from the Austrian Academy of Sciences ({\"O}AW).
\end{acknowledgments}

\bibliographystyle{apsrev4-1_abbrv}
\bibliography{literature}

\onecolumngrid
\clearpage

\renewcommand{\theequation}{S\arabic{equation}}
\renewcommand{\thefigure}{S\arabic{figure}}
\setcounter{equation}{0}
\setcounter{figure}{0}
\setcounter{page}{1}

\section{Supplementary Material for: "Organic polariton lasing and the weak- to strong-coupling crossover''}
\twocolumngrid

\section{Role of diamagnetic $A^2$ term}
\label{sec:role-diamagnetic-a}

This section discusses the role of the diamagnetic $A^2$ term, allowing one to disentangle the effects of this term vs other effects of strong coupling on the form of the phase diagram.

As noted in the text, this $A^2$ term is included in order to avoid a ground-state superradiant transition, following
the results of~\citet{rzazewski1975phase}.  Without the diamagnetic term, the ground state of the Dicke model at strong coupling is a state with a macroscopic occupation of the photon~\cite{hepp1973equilibrium}, even in
the absence of pumping.  The inclusion of the $A^2$ term, with a coefficient determined by the oscillator strength sum rule prevents this transition occurring~\cite{rzazewski1975phase}, so the ground state remains stable at all coupling strengths. Recently the question of whether a phase transition is in fact possible in the ground state has been re-opened, with a consensus developing that a transition is possible, but driven by dipole-dipole interactions between atoms when in the appropriate geometry~\cite{griesser16,de2018cavity}.  We neglect the possibility of this ferromagnetic transition here, and include an $A^2$ term so that only the transition induced by incoherent pumping is present.  

\begin{figure}[htpb]  
\includegraphics[width=0.99\linewidth]{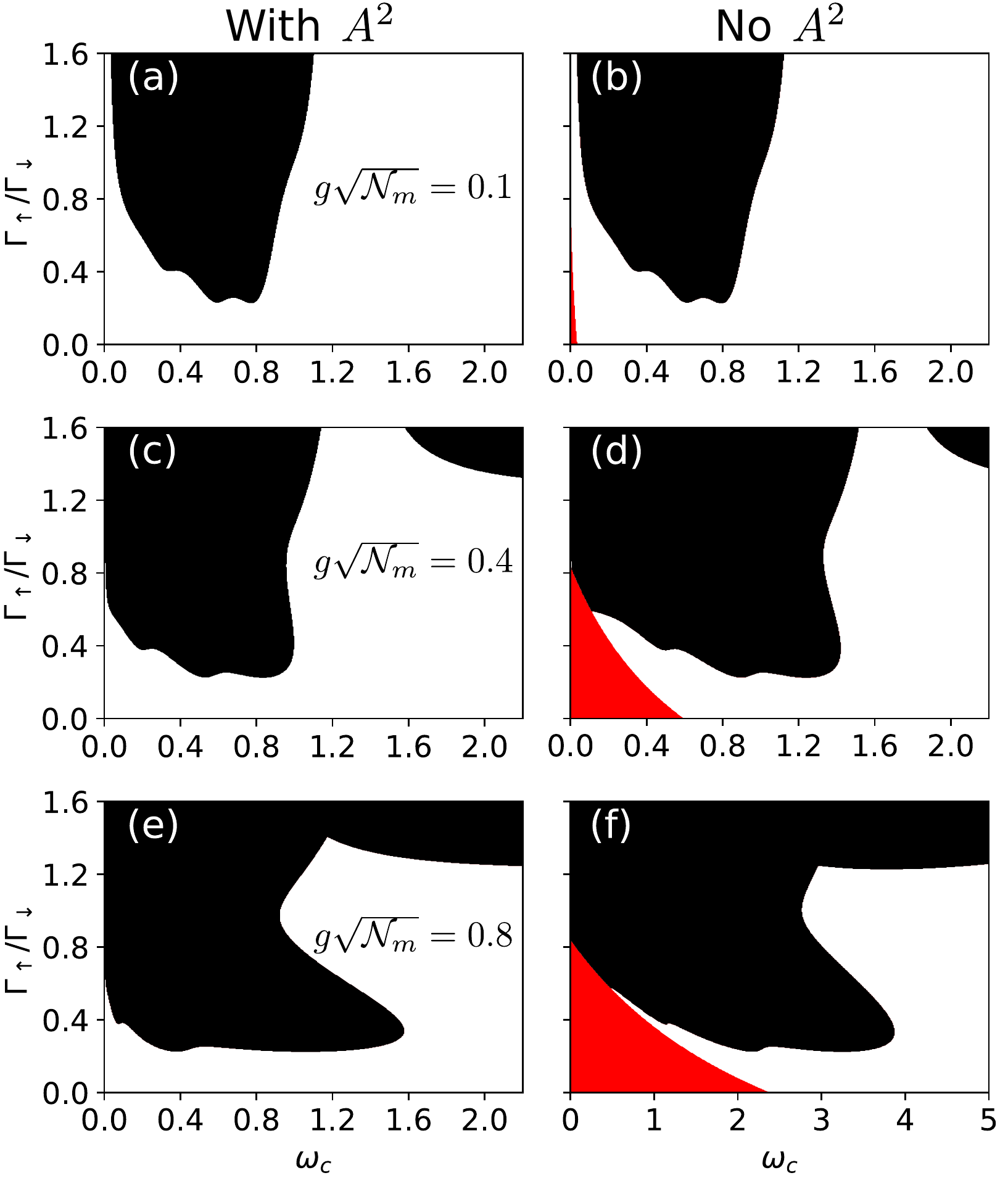}
\caption{Role of $A^2$ term, fixed coupling, parameters as in
  Fig.~\ref{cartoon}. White region denotes normal state, black denotes
  lasing, and red the superradiant phase.  Note the different $x$-axis
  scale in panel (f).}
\label{A2_term_effect__N4}
\end{figure}

Figure~\ref{A2_term_effect__N4} compares the phase diagrams (vs bare cavity
frequency and pumping) with (left) and without (right) the diamagnetic $A^2$
term.  In the phase diagrams without the $A^2$-term a superradiant phase is present at  low photon frequencies and low pumping.  As discussed elsewhere~\cite{kirton2018superradiant}, this superradiant state can be distinguished from the lasing state by the nature of the instability and the spectrum of the emission:  ground state superradiance is a stationary steady state, while the lasing state is at finite frequency, and thus is time dependent in the lab frame. As a consequence, the superradiant state instability involves a single unstable frequency (regions marked red on Fig.~\ref{A2_term_effect__N4}), while the lasing state instability involves a complex conjugate pair of unstable modes (regions marked black on Fig.~\ref{A2_term_effect__N4}).

Regarding the effect of the $A^2$ term on the shape of the polariton lasing region, we see from Fig.~\ref{A2_term_effect__N4} that the lasing phase remains a similar shape, although the characteristic frequency ranges are significantly rescaled at strong coupling. 
As pointed out in the Letter, the $A^2$ term has a twofold effect. Firstly, it leads to renormalization of the effective photon frequency $\omega_c^{\text{eff}}$, which leads to a $g$-dependent effective detuning. Secondly, it reduces the effective exciton-photon coupling. 
In the  ultrastrong coupling shown in Fig.~\ref{A2_term_effect__N4}(e,f),  the combination of these effects results in the effective decoupling of the photon from the molecular excitations, causing the lasing region to shrink to a significantly lower frequency range.
As the shape is similar, we can be clear the bending of the lasing lobe is not associated with the $A^2$ term.

\section{Effect of vibrational truncation}
\label{sec:effect-vibr-trunc}

\begin{figure}[htpb]  
\includegraphics[width=0.99\linewidth]{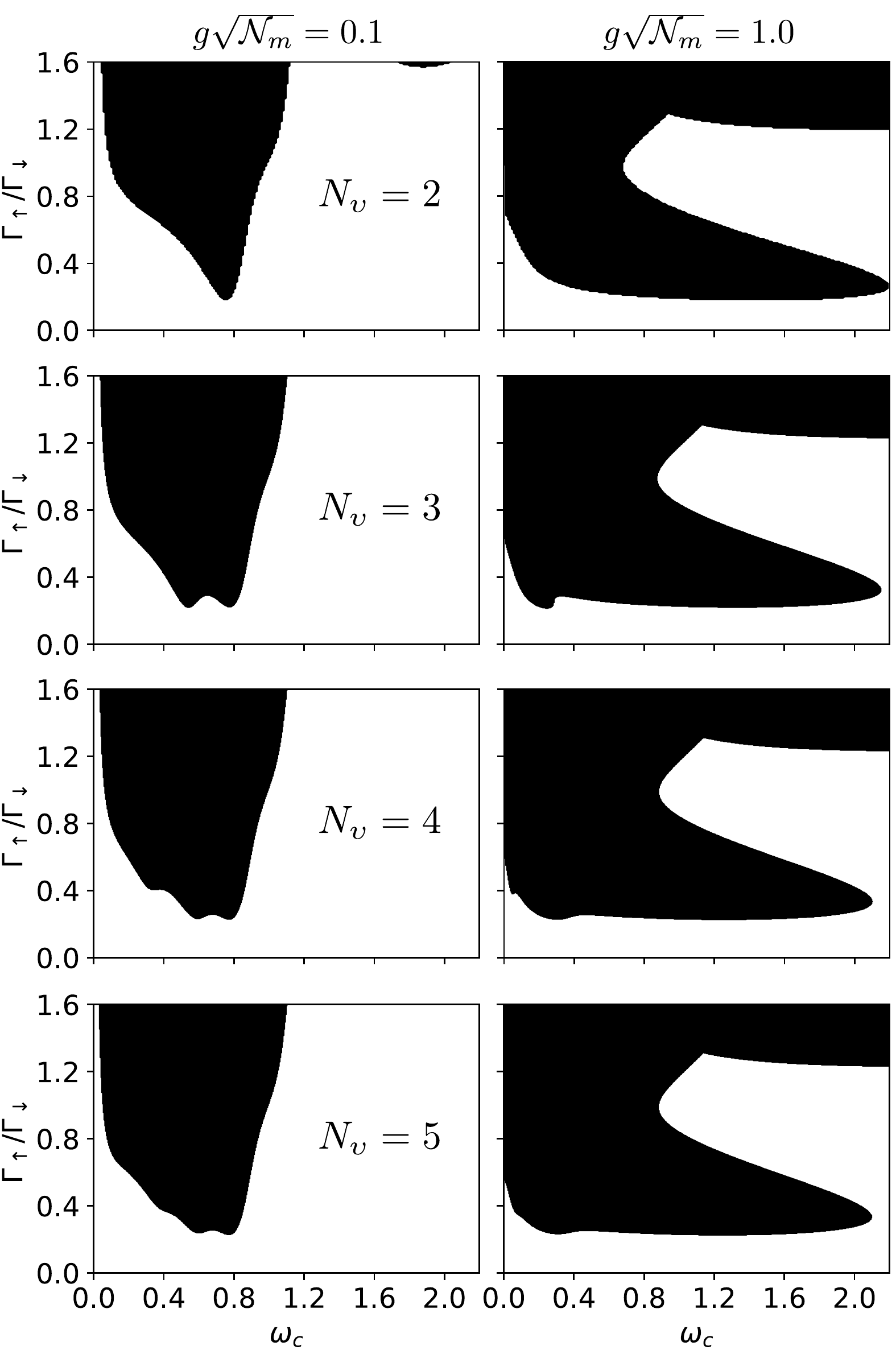}
\caption{Dependence of phase diagrams on the number of vibrational levels
  retained, $N_v$.  Other parameters as in Fig.~\ref{cartoon}.}
\label{effect_of_N}
\end{figure}

In Fig.~\ref{effect_of_N} we show the dependence of the phase diagram on the number of vibrational states $N$.  We show results for $S=0.1$ and for two values of matter-light coupling, $g \sqrt{\Nmol} = 0.1$ and $g \sqrt{\Nmol} = 1.0$, corresponding to the top left and bottom right panels of Fig.~\ref{ph_diagrams_g_goes_up} in which we used $N_v=4$. 
For the weaker coupling, $g \sqrt{\Nmol} = 0.1$, as $N_v$ increases the phase diagram retains a similar structure, but one can see the appearance of new peaks at $\omega_c \approx 0.6, 0.4, 0.2$ corresponding to (2-0), (3-0) and (4-0) transitions respectively.  For stronger coupling, $g \sqrt{\Nmol} = 1.0$, we see a similar emergence of extra peaks, and a slight evolution of the extended lasing lobe.  In both cases we see that the results do not differ significantly between $N_v=4$ and $N_v=5$, confirming we can safely use $N_v=4$ in our Letter. 

For $S=0.1$, we can in fact expect from simple arguments that the dominant physics  comes from single vibrational excitation. For an isolated molecule, neglecting strong coupling,  the probability for a ($n$--0) transition is $P_n = e^{-S} S^n / n!$, thus for $S=0.1$, we would expect results to be dominated by $n=0,1$.  The behavior shown in Fig.~\ref{effect_of_N} shows that despite this, higher vibrational states do play some role even for these small values
of $S$.

\section{Nature of the lasing instability}

In this section we provide further details on the calculation of the vibrational composition of the lasing mode, and present figures illustrating the evolution of the spectrum.

\subsection{Composition of linear stability eigenmodes}

To determine the molecular transitions corresponding to a given unstable mode, we note that from the eigenvector $\upsilon = \left( \delta \alpha , \delta \alpha^* , \mathbf{\delta\ell} \right)^{\intercal}$ of the stability
matrix $\mathcal{M}$, we can extract the matter part $\delta \ell_i$ and  so construct the corresponding molecular density matrix,
\begin{math}
  \rho(t) = \rho_{\text{ns}} + 
  \mathcal{A} (\delta \ell_i e^{\xi t} + \delta \ell_i^{\ast} e^{\xi^{\ast} t}) 
  \lambda_i/2
\end{math}
where $\mathcal{A}$ is an arbitrary amplitude and $\rho_{\text{ns}}$ is the normal state density matrix.  (For simplicity of notation, we neglect the superscripts on $\delta \ell$ and $\xi$ labeling eigenmodes.)
The complex conjugates appearing here are required in order to guarantee
Hermiticity (given the Gell Mann matrices are defined to be Hermitian). This step is crucial since the equations of motion mix $\ell_i$ and $\ell_i^\ast$.
To find the amplitude of the oscillatory component of a given 
element of the density matrix $\rho_{ij}$, we first define the matrix
$r = \delta \ell_i \lambda_i/2$ and then note the oscillatory
component of $\rho_{ij}$ takes the form $\delta \rho_{ij} = r_{ij} e^{\xi t} + r_{ji}^\ast 
e^{\xi^\ast t}$.  

While the diagonal components of a density matrix give the molecular state populations, nondiagonal ones correspond to coherences. In particular,
the block of the density matrix which is off-diagonal in terms of electronic states gives the weights of molecular transitions involved in lasing.
For the diagonal components of the density matrix we can immediately write
$\delta \rho_{ii} = 2|r_{ii}| \cos\big( \xi^{\prime\prime} t + \text{Arg} (r_{ii})\big) e^{\xi^\prime t}$ where $\xi^\prime, \xi^{\prime\prime}$ denote real and imaginary parts of the eigenvalue respectively.  For the off diagonal components, the behavior is more complex
as in general $|r_{ij}| \neq |r_{ji}|$;  we thus expect that the quantity
$\delta \rho_{ij}$ traces an elliptical spiral in the Argand plane,
$\delta \rho_{ij} = [r_A \cos(  \xi^{\prime\prime} t + \phi) 
+ i r_B \sin(  \xi^{\prime\prime} t + \phi)] e^{i \theta+\xi^\prime t} $.  One may readily show
that the semi-major and semi-minor axes of this ellipse are given by $r_{A,B} = |r_{ij}| \pm |r_{ji}|$, while $\phi,\theta=[\text{Arg}(r_{ij}) \pm \text{Arg}(r_{ji})]/2$.   Given this behavior, we define the amplitude of a given component by the semi-major axis.  The contribution of a given molecular transition $(n-m)$ to the lasing mode thus corresponds to the amplitude
$|r_{n\downarrow, m\uparrow}| + |r_{m\uparrow,n\downarrow}|$.

\subsection{Evolution of spectrum with coupling}
\label{sec:evol-spectr-with}

\begin{figure}[htpb]  
\includegraphics[width=0.99\linewidth]{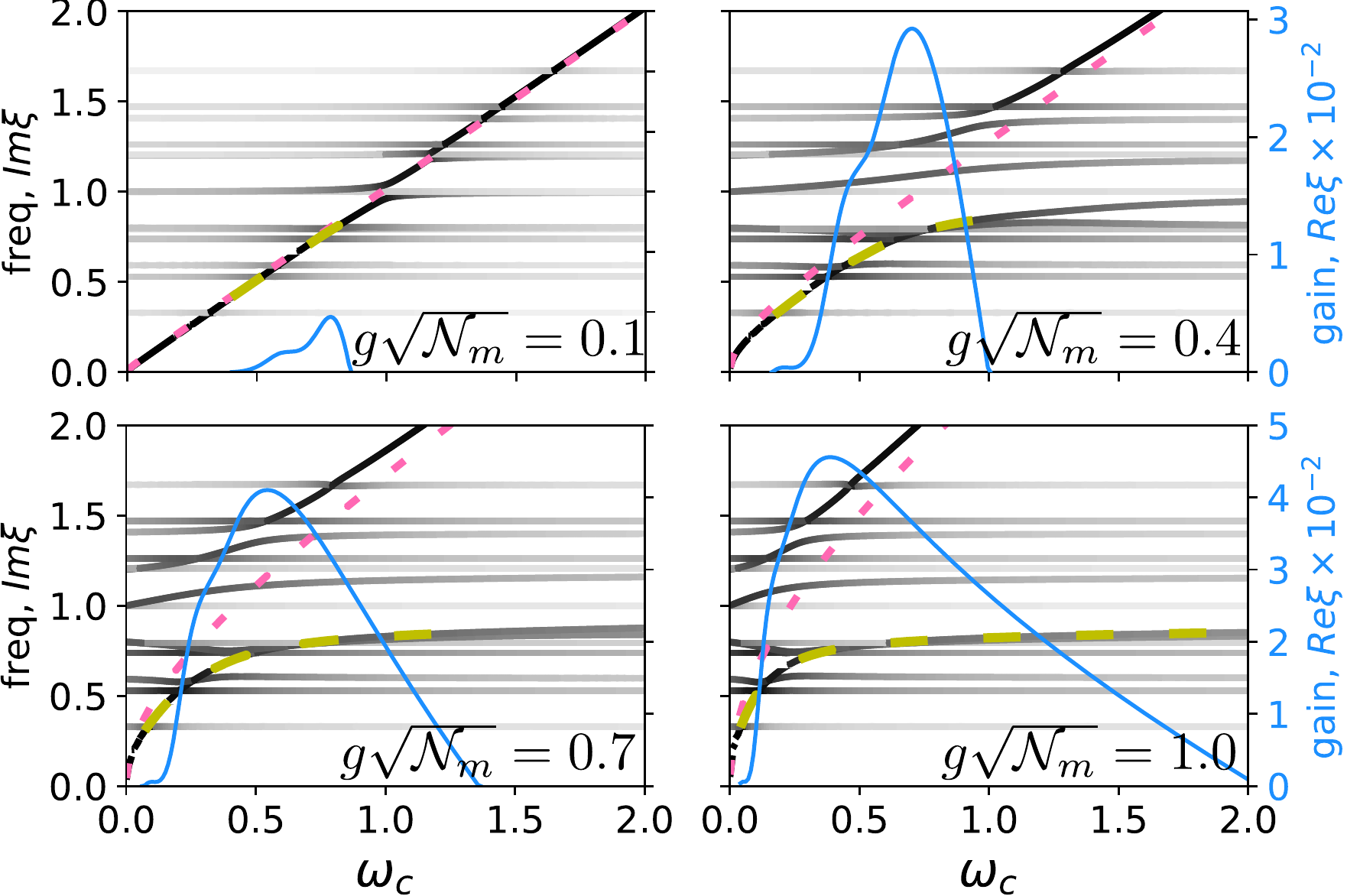}
\caption{Evolution of spectrum as coupling $g \sqrt{\Nmol}$ increases, for
  weak pumping $\Gamma_{\uparrow} = 0.4 \Gamma_{\downarrow}$.  Line colors
  and styles as in Fig.~\ref{spectrum_and_DM_cuts}, parameters as in
  Fig.~\ref{cartoon}.}
\label{spectrum_evolution_coupling_goes_up__N4}
\end{figure}

In Fig.~\ref{spectrum_and_DM_cuts} we showed the evolution of the spectrum with increasing pumping.  Here,  in Fig.~\ref{spectrum_evolution_coupling_goes_up__N4}, we complement this by showing how the spectrum evolves with increasing matter-light coupling at weak pumping.  We clearly see how the polariton splitting increases and self-tuning develops with increasing coupling. We may also note that the real part of the unstable eigenvalue increases, so the lasing instability will develop on a shorter timescale.

\subsection{Transition from polariton to photon lasing}
\label{sec:trans-from-polar}

\begin{figure}[htpb]  
\includegraphics[width=0.99\linewidth]{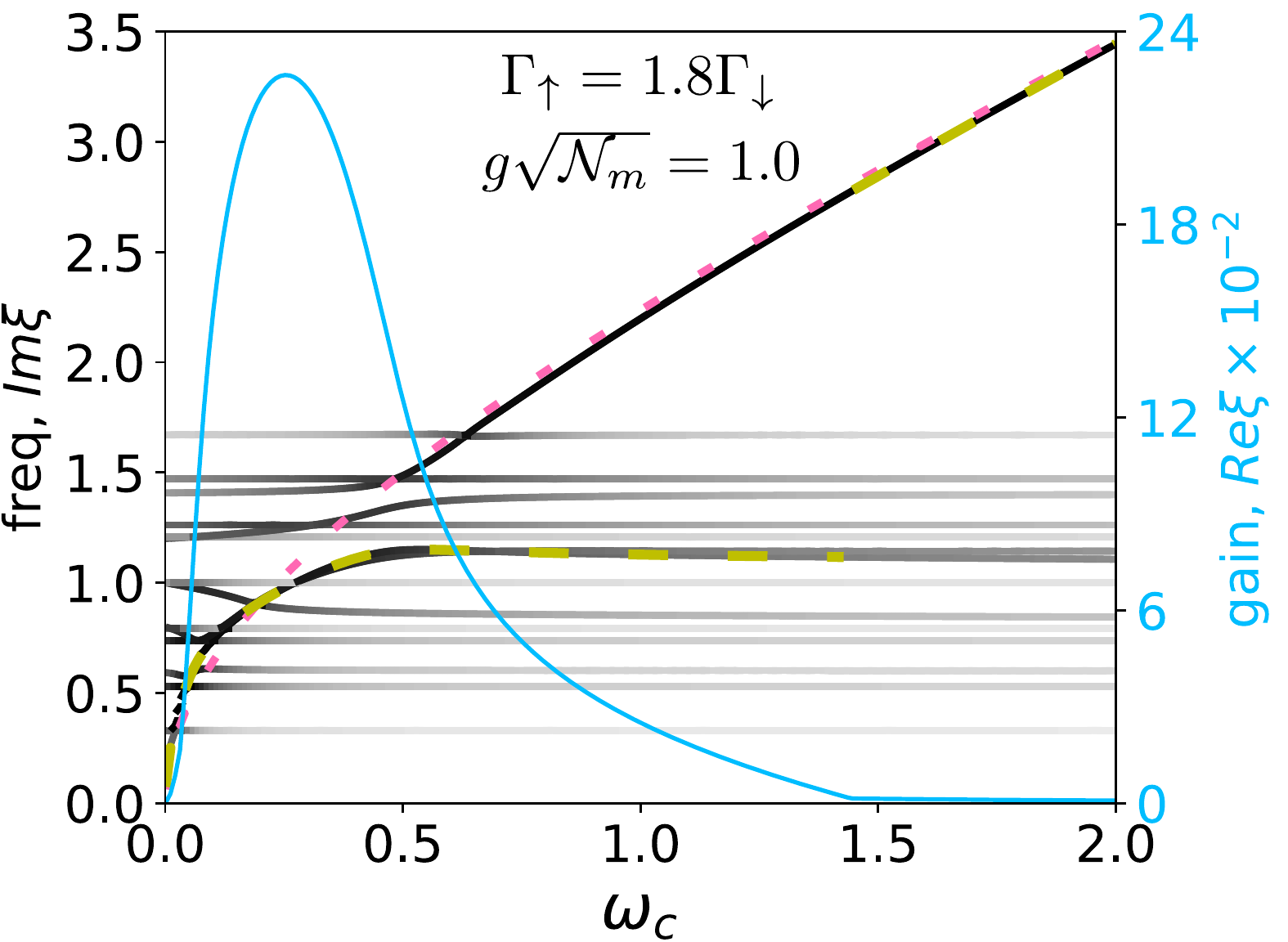}
\caption{Spectrum for $g \sqrt{\Nmol} = 1.0$, $\Gamma_{\uparrow} = 1.8
  \Gamma_{\downarrow}$ on an enlarged axis.  Line colors and styles as in
  Fig.~\ref{spectrum_and_DM_cuts}, parameters as in Fig.~\ref{cartoon}.}
\label{sp_pump_18__coupl_10}
\end{figure}

In Fig.~\ref{sp_pump_18__coupl_10} we show the spectrum for the same parameters as in Fig.~\ref{spectrum_and_DM_cuts}(i) (strong coupling and strong pumping), but over an extended axis range.This figure illustrates the transition between the two different lasing states which occurs in this regime. For low cavity frequencies (around $\omega_c \lesssim 1.5$) the system supports self-tuning and we get polariton lasing. 
However at larger cavity frequencies, the larger detuning in combination with strong pumping suppresses strong coupling leading to ``normal'' lasing at
effective photon frequency $\omega_c^{\text{eff}}$. As this regime leads to  effectively weak coupling photon lasing, we find the growth rate for this second lasing instability is much slower than in the polariton lasing regime.

We note that in this figure, because we use $N_v=4$, we see only a limited range of vibrational sidebands.  A calculation with large $N_v$ would show a sequence of sidebands up to higher frequency (but with a weight that decreases as one goes to higher frequencies).

\section{Phase boundary vs coupling strength}

In this section we discuss further the phase boundaries and minimum lasing threshold shown in Fig.~\ref{threshold}.  We provide further cuts of the phase diagram at different cavity frequencies, present the optimal cavity frequency at a given coupling strength, and provide details of the weak coupling calculation that is shown in Fig.~\ref{threshold}(a-c).

\subsection{Evolution of phase boundary with cavity frequency}
\label{sec:phase-boundaries-at}

\begin{figure}[htpb]  
\includegraphics[width=0.99\linewidth]{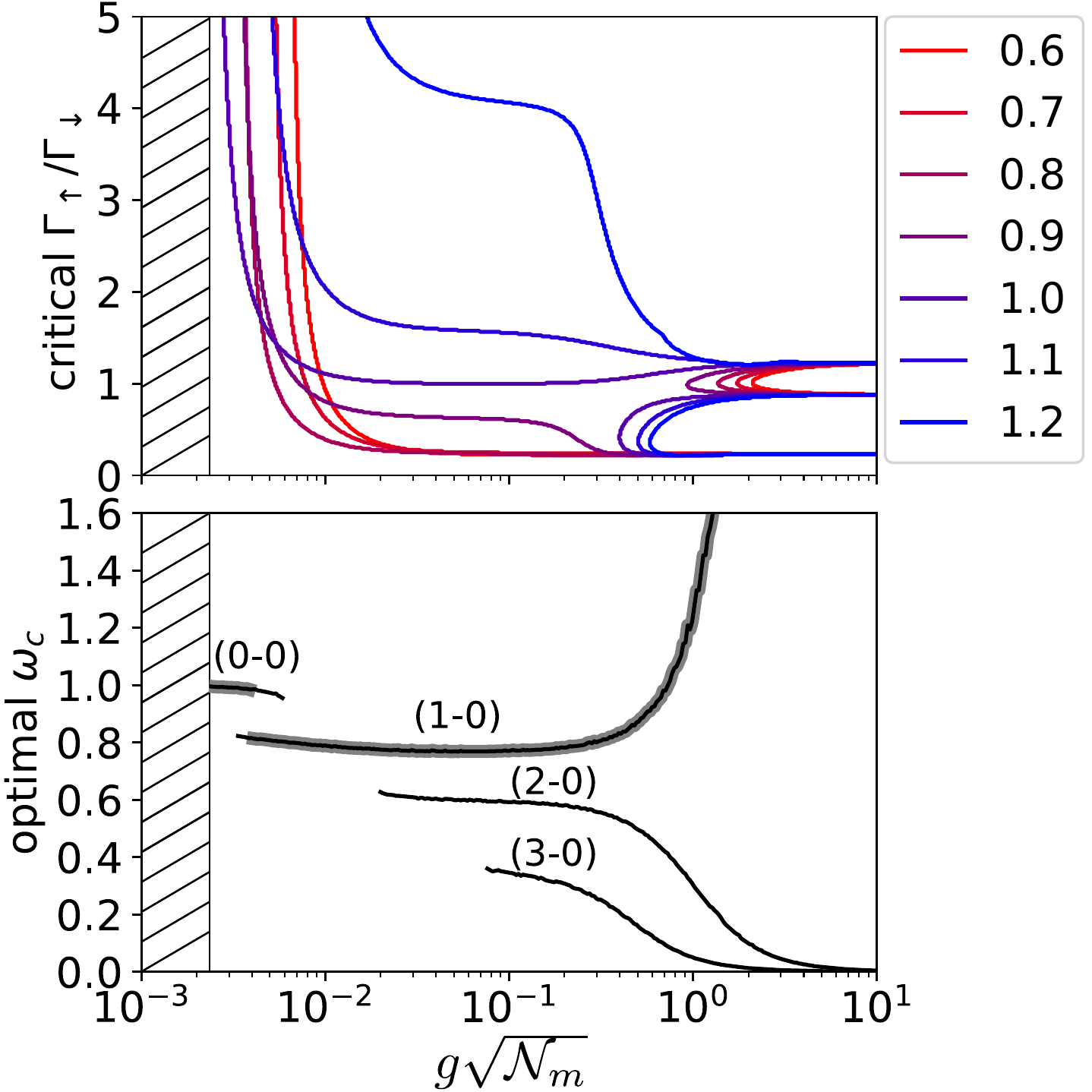}
\caption{Top: Critical pump strength as a function of coupling strength for
  various bare cavity frequencies $\omega_c$.  Bottom: The cavity
  frequencies $\omega_c$ at which the threshold has a local minimum as a
  function of coupling $g\sqrt{\Nmol}$.  The thick gray line highlights
  which frequency corresponds to the global minimum.  Parameters as in
  Fig.~\ref{cartoon}.}
\label{min_coupling_differnt_freq}
\end{figure}

We first provide further detail on the how the phase boundary evolves with cavity frequency.
Figure~\ref{threshold}(a-c) presents the phase diagrams at three specific frequencies; to show more clearly how the phase boundary evolves with frequency, the top panel of
Fig.~\ref{min_coupling_differnt_freq} shows a sequence of curves, giving the evolution of the phase boundary with bare photon frequency. Note that the 
phase boundary has different topology for $\omega_c \leq 0.9$ and $\omega_c \geq 1$.  For low frequency, there is a single lasing region (but with a hole 
appearing at large coupling around $\Gamma_\uparrow \simeq \Gamma_\downarrow$).  Around $\omega_c=1$, this hole reaches the phase boundary, dividing the 
lasing region into two: a large region at high pumping, and a smaller region, below inversion and at strong coupling.

\subsection{Optimal cavity frequency}
\label{sec:optim-cavity-freq}

The lower panel of Fig.~\ref{min_coupling_differnt_freq} shows the cavity
frequencies at which the threshold has local minima.  As could already be seen in 
Fig.~\ref{ph_diagrams_g_goes_up}, the threshold shows multiple local minima
at vibronic replicas $\omega_c = \varepsilon - n \omega_v$.  At the lowest
coupling, the minimal threshold is near $\omega_c = \varepsilon$ because
this corresponds to the peak emission --- for very low coupling, the
critical condition is for emission to beat cavity loss. Note that the
threshold lasing power diverges around $g\sqrt{\Nmol}\simeq 0.0024$,
i.e.\ for the parameters we use, there is no lasing at weaker coupling than
this. When coupling increases to $g \sqrt{\Nmol} \approx 0.0035$, we get
another local minimum emerging at $\omega_c \approx \varepsilon - \omega_v$
corresponding to the (1-0) transition.  For $g \sqrt{\Nmol} \gtrsim 0.004$
this new minimum becomes the global minimum, i.e.\ it leads to the lowest
threshold pumping as shown in Fig.~\ref{threshold}(d). As we enter strong
coupling regime $g \gtrsim 0.03$ we get one more local minimum driven by
(2-0) transition and then another one due to (3-0) transition. All these
local minima of pumping can be seen in Fig.~\ref{ph_diagrams_g_goes_up}.

Increasing the coupling further (into the strong coupling self-tuning
regime) we see a continuous evolution of (2-0) and (3-0) branches to lower
frequency, while the (1-0) branch moves to higher frequencies.  The
evolution of the (1-0) transition frequency can be associated with the
self-tuning effect which occurs at strong coupling, i.e.\ self-tuning allows
the lasing threshold for this transition to move out to high cavity frequencies.
In contrast, the (2-0) and (3-0) transitions do not show any
strong-coupling self-tuning effect --- the lasing transitions at these
frequencies appear similar to weak coupling lasing.  The reason these
lasing frequencies shift to lower frequencies is the renormalization of
effective photon frequency by the $A^2$ term.  i.e., the resonance condition is in fact $\omega_c^{\text{eff}} \simeq \varepsilon - n \omega_v$, and the dependence of $\omega_c^{\text{eff}}$ on $g\sqrt{\Nmol}$ means the resonance moves
to lower bare cavity frequencies as coupling increases.

In the ultrastrong coupling regime, the (1-0) and (2-0) branches give
almost the same threshold pumping around
$\Gamma_{\uparrow}^{th} \approx 0.22 \Gamma_{\uparrow}$, while the (3-0)
transition has $\Gamma_{\uparrow}^{th} \approx 0.38 \Gamma_{\uparrow}$,
as can also be seen in Fig.~\ref{ph_diagrams_g_goes_up}.

\subsection{Comparison to weak coupling theory}
\label{sec:comp-weak-coupl}

In Fig.~\ref{threshold}, we compared the strong coupling phase diagram to a weak coupling prediction of the phase boundary. Here we provide details of this weak coupling prediction.  The weak coupling theory of the Dicke-Holstein model has been discussed in Ref.~\cite{kirton2013nonequilibrium,phot_cond_theor_PRA}.  In those works, it is shown that treating matter-light coupling perturbatively, one can derive a weak coupling master equation:
\begin{multline}
  \dot{\rho}(t) = 
  -i \left[ \tilde{H}_0, \rho \right] + \kappa \mathcal{L}[a] + 
  \sum_n \Big(
  \Gamma_{\downarrow}  \mathcal{L}[\sigma_n^{-}] +
  \Gamma_{\uparrow}  \mathcal{L}[\sigma_n^{+}] 
  \\  +
  \Gamma_E \mathcal{L}[a^{\dagger} \sigma^-] + \Gamma_A \mathcal{L}[a \sigma^+]
  \Big)\rho,
\end{multline}
where the free Hamiltonian is  $\tilde{H}_0 = (\omega_c - \varepsilon) a^{\dagger} a$ and the rates of emission and absorption processes are given by 
\begin{equation}
  \Gamma_{A,E}  =  g^2 \int_{-\infty}^{\infty} dt e^{\pm i \omega_c t}  \langle \sigma^-(t) \sigma^+(0) \rangle_{0}.
\end{equation}
The two-time correlation function $\langle \ldots \rangle_0$ is calculated for free molecules, neglecting the matter-light coupling.
In Refs.~\cite{kirton2013nonequilibrium,phot_cond_theor_PRA}, a different form of dissipation was assumed for the vibrational modes, allowing an explicit form for $\Gamma_{A,E}$ through a Keldysh path integral.  The form of dissipation considered here and written in Eq.~(\ref{eq:orig_lindblad}) does not allow such a form, however the integrals in $\Gamma_{A,E}$ can be numerically calculated from the master equation using the quantum regression theorem~\cite{Breuer2002}.

From this equation, we can use the same mean-field decoupling discussed earlier, i.e.\ $\langle a \sigma^{\pm} \rangle = \langle a \rangle \langle \sigma^{\pm} \rangle $ and 
$\langle a^{\dagger} a \rangle = |\langle a \rangle|^2$ to yield a set of non-linear coupled equations:
\begin{align}
\partial_t \langle a \rangle =&  
    \frac {\Nmol} 4 \Big[ 
    \Gamma_A\left(1- \langle \sigma^z \rangle\right)
    - \Gamma_E \left(1+ \langle \sigma^z \rangle\right)
     \Big] \langle a \rangle
    \nonumber\\
    &
    - \left[ i (\omega_c - \varepsilon) + \frac{\kappa}{2}\right]
    \langle a \rangle,
  \\ 
\partial_t \langle \sigma^z \rangle =& 
    \Big( 1 - \langle \sigma^z \rangle \Big) \Big( \Gamma_{\uparrow} + \Gamma_A |\langle a \rangle|^2 \Big) \nonumber\\&- 
    \Big( 1 + \langle \sigma^z \rangle \Big) \Big( \Gamma_{\downarrow} + \Gamma_E |\langle a \rangle|^2 \Big).
\end{align}
We can then again consider linear stability of the normal state, by considering fluctuations $\langle a \rangle = \alpha$ and $\langle \sigma^z \rangle = \langle \sigma^z \rangle_{ns} + z$
as in the Letter. One may see that $\alpha$ and $z$ are uncoupled, so the corresponding equations can be solved trivially, and the linearized  eigenvalue  is given by
\begin{equation}
\xi =  - i(\omega_c-\epsilon) - \frac{\kappa}{2} + \frac{\Nmol}4 (\Gamma_E-\Gamma_A) + \frac {\Nmol} 4 (\Gamma_E+\Gamma_A) \langle \sigma^z \rangle_{\text{ns}},
\end{equation}
where $\langle \sigma^z \rangle_{\text{ns}} = (\Gamma_{\uparrow} - \Gamma_{\downarrow}) / (\Gamma_{\uparrow} + \Gamma_{\downarrow})$. 
One can extract threshold pumping as a function of coupling $g$, 
by solving $\text{Re}[\xi] = 0$, which gives:
\begin{equation}
\label{gam_th_from_az}
\Gamma_{\uparrow}^{th} = \frac{\Nmol\Gamma_A + \kappa}{\Nmol\Gamma_E - \kappa} \Gamma_{\downarrow}.
\end{equation}

\end{document}